# Develop Health Monitoring and Management System to Track Health Condition and Nutrient Balance for School Students

Mohammad Ali
Computer Science and Engineering
United International University
Dhaka, Bangladesh
mali122013@bscse.uiu.ac.bd

*Abstract*—Health Monitoring and Management System (HMMS) is an emerging technology for decades. Researchers are working on this field to track health conditions for different users. Researchers emphasize tracking health conditions from an early stage to the human body. Therefore, different research works have been conducted to establish HMMS in schools. Researchers propose different frameworks and technologies for their HMMS to check students' health condition. In this paper, we introduce a complete and scalable HMMS to track health conditions and nutrient balance for students from primary school. We define procedures step by step to establish a robust HMMS where big data methodologies can be used for further prediction for diseases.

*Keywords—Health Monitoring and Management System, Student Health Condition, Student Nutrient Balance, Big Data*

## I. INTRODUCTION

Healthcare is one of the basic needs of a human being. Therefore, basic health education has been introduced in every educational curriculum. Students receive sufficient health support from their educational institutes like schools as well as new technologies have been introduced in the health management system to ensure sustainability. Researchers propose an automatic health screening system that takes some basic key factors such as height, weight, blood pressure, etc. and store those data for the specific student. They use the electronic card to uniquely identify the specific students' health data [1]. In another research, wireless technology has been used to develop a health monitoring system for kindergarten students which uses some basic parameters such as body temperature, heart pulse rate, etc. [2].

Health data creates an extremely large dataset that is difficult to process with a typical database management system but big data analytics may help to resolve the problem. On another part, data mining and pattern recognition technology may be used on big data to find which treatments are more effective for a specific health condition, identify drug side effects, identify prior disorders that can help patients, and reduce cost [3]. Researchers also emphasize the privacy and security of data to avoid unauthorized access from the other side of the system. Cloud technology [4], wearable sensors, mobile phone [5] have also been using on personal health record which allows to create, manage, and control health information.

In this paper, we propose a complete, scalable health monitoring and management system (HMMS) to track health conditions and nutrient balance for students. We take student's health data during admission time, analyze those data, generate the result by using clinical standards, and refer students to the doctor if the generated result has any health issues. We start our work by defining parameters for health screening, then we design the system and define different accessibility for different users. We conclude our work with a discussion of how big data technologies can be introduced with our system in the future.

## II. RELATED WORK

The health management system or health monitoring system is not a new innovation. Many research works have been conducted to build robust health management and monitoring system. Researchers proposed a wireless patient monitoring system for the hospital to monitor patient activities like sleeping, sitting, breakfast/lunch/dinner habits, medication, etc. Their proposed system collects this activity information, analyze it, and detect the abnormal behavior of the patient [6][7]. Some research works have been conducted for specific patients' or groups, for example, Yoke et al. developed a web and mobile application for pregnant mother and their children. This application provides relevant health information, early detection problems, and a reminder of important events for pregnant mother and her children [8]. In another work, researchers proposed a mobile-based healthcare application that takes physical input parameters, analysis, and suggest diet instruction [9]. We found a group of researches for diabetic patients. Researchers propose a self-managing system for diabetes patients by using gamification and social networking platform. The system helps patients to learn their diabetes condition in an interactive way [10][11][12]. Researchers describe different features of the health monitoring system. These features build relationships between health data and showed how data can be managed usefully [13]. They also emphasize the impacts of health information technology on healthcare quality and safety. They showed that their system has positive effects which could give assistance to clinical decision making and enhance patient satisfaction [14].

## III. SYSTEM DEVELOPMENT

### A. System Design

Students will get an RFID (Radio Frequency IDentification) card after getting admission into the school. School authority will organize a health camp every year after admission. Nurses will do different kinds of health checkup for students and collect health data and input into our system. Our system will analyze those data and generates results using the clinical standard and store it in the database. If any student's data fail to any clinical standard, then it will refer the student to the doctor. Doctor punch student's RFID card in the system to get all data for the particular student and give the student proper treatment. Fig 1, shows the system design of our proposed system.

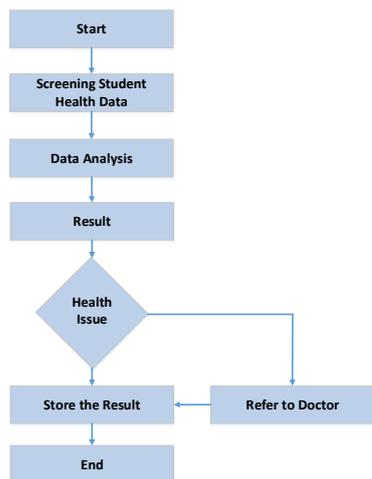

Fig. 1. System Design of Our Proposed System

### B. Identify Parameters

We took interview four child doctors to know about the parameters to check student's health conditions and nutrition balance. The age range for students usually from 4 years old to 16 years old. We have shorted out 45 (N = 45) parameters for that age range of students. We categorized those parameters into two groups: One-time entry parameters and multiple time entry parameters. Table I, shows one-time entry and table II, shows multiple time entry parameters.

TABLE I. ONE-TIME ENTRY PARAMETERS/FEATURES

| Parameters | Area |
|---|---|
| Student Name | General Information |
| Student Photo | General Information |
| Screening ID | General Information |
| Student School ID | General Information |
| Date of Admission in the School | General Information |
| Birth Certificate Number | General Information |
| Date of birth | General Information |
| Place of birth | General Information |
| Father's Name | General Information |
| Mother's Name | General Information |
| Address | General Information |
| Family History of any Disease | General Information |
| Vaccination Status | Vaccination |
| Blood Group & RH Typing | Clinical Test |
| Birth Weight | General Information |
| Childbirth History | General Information |
| Breast Feeding Status | General Information |
| Teeth – When it grown | Dental Condition |

TABLE II. MULTIPLE-TIME ENTRY PARAMETERS/FEATURES

| Factor/Attribute | Area |
|---|---|
| Present Class | General Information |
| Height | General Information |
| Weight | General Information |
| Helminthiasis Status | Nutrition |
| Vision Condition | Eye Condition |
| Night Blindness | Eye Condition |
| Hearing | Hearing Condition |
| Dental Caries | Dental Condition |
| Tonsil Condition | ENT |
| MUAC / MAC | Clinical Test |
| H/O - Recent illness | General Information |
| BMI – Body Mass Index | General Information |
| PEM - Protein Energy Malnutrition | Nutrition |
| Skin Problem/ Skin Disease | Skin Condition |
| Iodine [IQ Test] | Mental Condition |
| CBC and ESR | Clinical Test |
| HbsAg | Clinical Test |
| Urine R/E | Clinical Test |
| Stool R/E | Clinical Test |
| TSH (Thyroid Stimulating Hormone) | Clinical Test |
| Food Taste Condition | General Information |
| Complementary Food Given | General Information |
| Nail Condition | Hygienic Information |
| Junk Food/ Fat Food Habit | General Information |
| Behaver | Mental Condition |
| History of Asthma | General Information |
| Nose polyps | ENT |

### C. Identify Clinical Standard

After identifying the parameters, we identify the clinical standard for those. For example, we have to ensure that every student must take all vaccination doses provided by the Bangladesh government listed in Table III.

TABLE III. VACCINATION/IMMUNOLOGY TABLE [15]

| Vaccine | Disease Prevented | Number of Doses | Recommended Age for Doses |
|---|---|---|---|
| BCG | Tuberculosis | 1 | After Birth |
| Pentavalent Vaccine | Diphtheria, Pertussis, Whooping cough, Haemophilus Influenzae Type B | 3 | 6, 10, 14 Weeks |
| PCV | Pneumococcal disease | 3 | 6, 10, 14 Weeks |
| OPV (Polio) | Poliomyelitis (Polio) | 3 | 6, 10, 14 Weeks |
| IPV | | 2 | 6, 14 Weeks |
| MR Vaccine 1 | Measles and Rubella | 1 | 9 Months |
| MR Vaccine 2 | | 1 | 15 Months |

### D. Define Users

We define four users for our system with different accessibility. Those are Admin, Nurse, Doctor and Patient/Student/Parents.

TABLE IV. DIFFERENT TYPES OF USERS WITH ACCESSIBILITY

| Users | Accessibility |
|---|---|
| Admin | • Create, update, delete, view doctor and nurse information<br>• View students basic and health information<br>• Delete student's health information |
| Nurse | • View student's information by<br>   o Searching student id (Unique)<br>   o Punching RFID ID<br>• Input health data for specific student<br>• Edit, view, print student's health data |
| Doctor | • View student's information by<br>   o Searching student id (Unique)<br>   o Punching RFID ID<br>• View, print student's health data<br>• Recent health data<br>• Old health data |
| Student/Patient/Parents | • View minimal/basic information about their health information<br>• Notice/suggestion/references |

Figure 2, shows the web features for admin and Figure 3, shows web features for nurse and doctor.

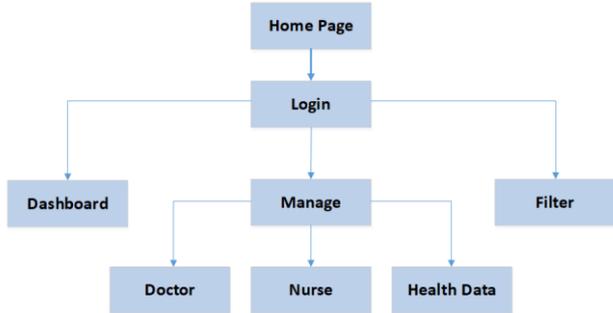

Fig. 2. Accessibility diagram for Admin

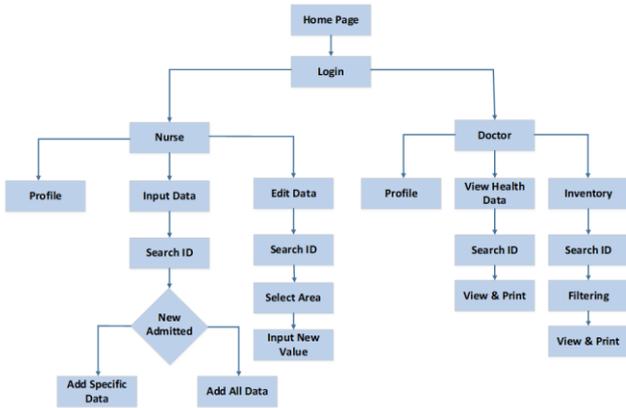

Fig. 3. Accessibility diagram for Nurse and Doctor

## IV. OUTCOME AND FUTURE WORK

So far, we have designed the system, identified the parameters, identified the clinical standard, and define users with different accessibilities. This is an ongoing research work, so we have to work more on the next steps. Our next work is to make a user experience analysis, design, and develop the system. After developing the system, we have to input real health data into the system. We have to ensure the testing, verification, and validation of data. After successfully developing the system, we may identify diseases mentioned in table V and we may also introduce big data technology on those data on specific prediction targets. We can easily construct cohort and feature to make a predictive model and finally, we can evaluate the performance of the system by using big data techniques.

TABLE V. TEST LIST TO IDENTIFY DISEASE [16]

| Test Name | Disease |
|---|---|
| CBC | Anemia, infection, inflammation, bleeding disorder or cancer |
| ESR | To detect the presence of inflammation caused by one or more conditions such as infections, tumors or autoimmune diseases; to help diagnose and monitor specific conditions such as temporal arteritis, systemic vasculitis, polymyalgia rheumatica, or rheumatoid arthritis |
| Blood Group & RH Typing | To determine your ABO blood group and Rh type |
| HBsAg | Primarily to screen for and diagnose acute or chronic hepatitis B virus (HBV) infection, to detect a previous, resolved hepatitis B infection, or sometimes to guide and monitor treatment |
| Urine R/E | To screen for, help diagnose and/or monitor several diseases and conditions, such as kidney disorders or urinary tract infections (UTIs) |
| Urine R/M | The urinalysis is used as a diagnostic tool to identify different metabolic and kidney disorders. It is used to detect urinary tract infections (UTIs) and other disorders of the urinary tract. Patients with acute or chronic conditions, such as kidney disease, the urinalysis may be instructed at intervals as a rapid method to help record organ function, status, and response to treatment. Abnormal findings in urinalysis are a warning that something may be wrong and should be evaluated further. |
| Stool R/E | To determine whether you have an infection of your digestive tract due to the presence of disease-causing (pathogenic) bacteria |
| THS | To screen for and help diagnose thyroid disorders; to monitor treatment of hypothyroidism and hyperthyroidism |